\pdfoutput=1
\documentclass{emulateapj}
\usepackage{amssymb,amsmath,latexsym,natbib,wasysym}

\DeclareMathAlphabet{\mathpzc}{OT1}{pzc}{m}{it}

\newcommand\be{\begin{equation}}
\newcommand\ee{\end{equation}}

\begin{document}

\pagenumbering{arabic}

\slugcomment{Submitted to Astrophysical Journal, \today}
\shorttitle{Multiple Transiting Planets}
\shortauthors{Ragozzine \& Holman}

\title{The Value of Systems with Multiple Transiting Planets}

\author{Darin Ragozzine \& Matthew J. Holman}
\affil{Institute for Theory and Computation, Smithsonian Astrophysical Observatory, Cambridge, MA 02138}

\email{dragozzine@cfa.harvard.edu}

\begin{abstract}

Among other things, studies of the formation and evolution of planetary systems currently draw on two important observational resources: the precise characterization available for planets that transit their parent stars and the frequency and nature of systems with multiple planets. Thus far, the study of transiting exoplanets has focused almost exclusively on systems with only one planet, except for considering the influence of additional planets on the transit light curve, mostly through transit timing variations (TTVs). This work considers systems where multiple planets are seen to transit the same star and concludes that such "multi-transiting" systems will be the most information-rich planetary systems besides our own solar system. Five new candidate multi-transiting systems from \emph{Kepler} have been announced in Steffen et al. 2010, though these candidates have not yet been fully confirmed as planets. In anticipation of the likely confirmation of multi-transiting systems, we discuss the value of these systems in detail. For example, proper interpretation of transit timing variations is significantly improved in multi-transiting systems. The true mutual inclination, a valuable probe of planetary formation, can also be well determined in certain systems, especially through Rossiter-McLaughlin measurements of each planet. In addition, such systems may undergo predictable and observable mutual events, where one planet crosses over the other, which allow for unique constraints on various physical and orbital parameters, particularly the mutual inclination.

\end{abstract}


\maketitle

\section{Introduction}
\label{intro}

Even as transiting extra-solar planets are the most information-rich planets orbiting other stars \citep{2007prpl.conf..701C,2009IAUS..253...99W}, systems where multiple planets transit
their parent star will be the most information rich planetary \emph{systems} around main sequence stars after our own solar system. Though rarer than systems where one or no planets are seen to transit, these systems provide an exciting opportunity to learn more about the formation and evolution of planetary systems.

As of May 2010 there 
were no published examples of systems with more than one planet known to transit. In the last year, two transiting
planets were shown to have an additional companion well-characterized from radial velocity observations: the super-Earth CoRoT-7b \citep{2009A&A...506..287L,2009A&A...506..303Q} and the
hot Jupiter HAT-P-13b \citep{2009ApJ...707..446B}. HAT-P-13c has a much larger orbit than planet b, and it is not known whether it transits. CoRoT-7c is
in a 3.6-day period orbit, which is not transiting. Both of these systems are indicative of the trend that hot Jupiters tend to be the only massive object close to the star, while hot Neptunes and super-Earths show the opposite trend, tending to cluster in multiple planetary systems \citep[e.g.,][]{2010A&A...512A..48L} which has been predicted theoretically \citep{2007ApJ...654.1110T}. Though this inference is subject to observational bias, there are many examples from radial velocity surveys of close-in multi-Neptune systems with near circular orbits, which are quite likely to have low relative inclinations as well: coplanar analogues of the 61 Vir and HD 40307 systems that are inclined within 0.6$^{\circ}$ and 1.5$^{\circ}$ degrees of the line-of-sight, respectively, would  show all three planets transiting; HD 40307b is known not to transit \citep{2010arXiv1002.4707G}. 

The last decade of radial velocity observations has shown that the best way to find planets is to look for additional planets around stars with known planets. This applies to transiting planets as well: systems with transiting planets are much more likely to contain additional transiting planets than random stars \citep{2010arXiv1002.4702G,2010ApJ...712.1433B}. This is an extension of the early realization that if stellar binaries contained planets in the same plane, then eclipsing binaries are an ideal place to search for transiting planets \citep[e.g.,][]{1984Icar...58..121B,1990A&A...232..251S}. Several groups have searched for additional transiting planets around systems with known planets, including MOST \citep[e.g.,][]{2007ApJ...658.1328C}, EPOXI \citep{2010ApJ...716.1047B}, and Super-WASP \citep{2009MNRAS.398.1827S}. Unfortunately, these surveys have limited observations and statistical analyses show that they are seldom sensitive to planets with periods longer than $\sim$10 days, implying that only very close additional companions would be detectable. 

Observations by \emph{Kepler} have ushered in a new era of exoplanet studies, with the recent announcement of hundreds of candidate transiting exoplanets in \citet{2010arXiv1006.2799B}, with an estimated false positive rate that implies that in the first 43 days of operations, \emph{Kepler} discovered about as many new exoplanets than all previous methods combined over the last 20 years! Furthermore, this announcement was accompanied by \citet[][hereafter S10]{2010arXiv1006.2763S} with describe in detail 5 new systems where more than one planet candidate is seen to transit. Though these have not been confirmed as planets and some of these candidates could be false positives (see $\S$\ref{falsepos}), they demonstrate the ability of \emph{Kepler} to identify and characterize multi-transiting systems. As expected, the systems reported are mostly composed of smaller planets in long-period orbits. Most of the systems appear to be near mean-motion resonances and one system has three planetary candidates. It is important to note that the 5 systems reported are only a subset of multi-transiting systems identified by \emph{Kepler} to date (S10).

With their high sensitivity and long observing baselines, the \emph{CoRoT} mission \citep{2006cosp...36.3749B} and the \emph{Kepler Space Telescope} \citep{2010Sci...327..977B} are far more likely to find a system with two or more planets detected in transit around the same star. The lack of additional transiting planets around the reported \emph{CoRoT} systems seems to indicate that multi-transiting systems are only $\lesssim$10\% as common as transiting planets, in accordance with previous estimates \citep{2005Sci...307.1288H,2009IAUS..253..173F,2010EAS....42...39H}. Without full disclosure of the \emph{Kepler} population, the observed frequency of \emph{Kepler} multi-transiting systems is not clear, though they definitely appear to contain more than $\sim$1\% of the detected planetary candidates \citep{2010arXiv1006.2799B}. 

In this work, our goal is to focus on understanding systems in which multiple planets are already known to transit and for which there will be significant signal-to-noise in the transit light curves of both planets, such as the systems presented by S10. We focus on further anticipated results from the \emph{Kepler}, which is currently returning ultra-precise light curves from over 150000 stars \citep{2010ApJ...713L..79K}. That \emph{Kepler} would be likely to find and interpret such systems was anticipated as early as \citet{1996chz..conf..229K}.

This paper is a systematic investigation of the value of multi-transiting systems. In $\S$\ref{back}, we define our terminology, review the study of multiple planets around other stars, and consider the geometric probability of multi-transiting systems. Specific aspects of the value of multi-transiting systems in discussed in $\S$\ref{valuemts}. In $\S$\ref{model}, we describe a full numerical-photometric model that can produce self-consistent light curves of multi-transiting systems. Additional methods for studying multi-transiting systems, including mutual events and Rossiter-McLaughlin, are discussed in $\S$\ref{othersec}. We summarize with specific conclusions in $\S$\ref{concl}.

Much of the following discussion is relevant not only for systems with two planets in different orbits, but could also be directly or indirectly applied to systems with satellites (exomoons), trojans, transiting planets in eclipsing binary systems and ``multi-eclipsing'' stellar triples. The photometric effects of some of these have been modeled by other authors.

\section{Background}
\label{back}
\subsection{Terminology}
\label{term}
Systems where more than one planet is seen in transit do not yet have a standard terminology, so we find it valuable 
to first define some terms.\footnote{We also note that there is no standard term for a triple star system where 
eclipses between more than one pair are occurring; eclipsing triples can refer to a regular eclipsing binary with a 
non-eclipsing third star. By extension with the terminology we use, these could be called multi-eclipsing triples.} 
Multi-planet systems are stars orbited by more than one planet, whether or not any are seen in transit. A 
multi-transiting system (MTS) is one in which more than one of these planets (currently) transits the parent star and 
these are the subject of this paper. This can be made more specific, e.g. double-transiting system, triple-transiting 
system, etc. In most cases, not all of the planets will transit, so you can readily have a double-transiting triple 
system or a single-transiting quadruple system, etc. A mutual event refers to an overlap between two of the planets as 
seen from Earth in general and the terminology we use for these events (e.g., double transits, etc.) is described in 
$\S$\ref{mutev}.

\subsection{Transiting Planets with Companions}
\label{ttvback}
Most of the literature dealing with multi-planet transiting systems is focused on a transiting planet perturbed by an 
unseen second planet causing non-linear variations in the observed central transit time that cannot be explained by a 
purely periodic Keplerian orbit \citep[e.g.,][]{2002ApJ...564.1019M,2005MNRAS.359..567A,2005Sci...307.1288H}. Since transit times can with good observations be 
measured to 
within a few seconds\footnote{Measuring central transit times to within a few seconds is comparable to measuring 
orbital phase position to within 1 arcsecond.}, in the best cases, the presence or lack of transit timing variations 
(TTVs) is a powerful constraint on any non-Keplerian effect, such as an additional planet.

Though transit timing anomalies can be very large, as much as several minutes in the case of strongly resonant 
systems\footnote{Note that near-resonant systems also have large TTVs (Veras \& Ford 2010, submitted) and that the apparent period of a 
resonant planet (during a small fraction of the libration cycle) can be different from the true period by a couple percent, so the ratio of the apparent periods is not the only distinguishing factor.}, no announced planets 
have been detected using TTVs. The systems in S10 only cover 43 days of data and no significant TTVs are reported. A Monte Carlo analysis of expected TTVs over the course of the \emph{Kepler} mission indicates that these systems, especially the near-resonant ones, could show strong TTVs in the future (S10). We note that if these systems are nearly coplanar, as suggested by geometric arguments ($\S$\ref{probmulti}), the eccentricities estimated by S10 may also be low, as inclinations and eccentricities are usually dynamically coupled, suggesting that the actual TTV signals will be on the low end of the expectations. 

At the current time, there are a few hints of possible variability due to unseen 
companions in the literature \citep[e.g.,][]{2010arXiv1006.1348M}, but a strong clear TTV signal has yet to be demonstrated. Furthermore, the inversion 
process of taking transit times and inferring the properties of the perturbing planet appears is quite difficult in 
general (Veras \& Ford 2010, submitted), though not impossible \citep{2010ApJ...709L..44N,2008ApJ...688..636N,2010arXiv1005.5396M}. Besides a computationally difficult inversion problem, the TTV signal is often degenerate, with multiple possible configurations that each fit the data, even in systems where TTVs are measured quite accurately \citep{2007ApJ...664L..51F,2010arXiv1005.5396M}. This can be complicated even further when allowing for more than one perturbing planet (S10).

Considering systems where both planets are seen to transit helps the inversion problem immensely by constraining the period, phase, and size of the perturbing planet directly, without degeneracy. These parameters are enough to get good estimates on the perturbing planet's semi-major axis and mass, the primary determinants of the TTV signal \citep{2005MNRAS.359..567A,2005Sci...307.1288H}. Furthermore, as both planets perturb each other, there can be two sets of interdependent TTVs that must be 
self-consistent, which is particularly helpful in ruling out additional unseen perturbers (S10). We discuss the excellent value of such systems for constraining all the planetary system parameters in 
more detail below. Of course, using TTVs alone will still be needed in the case where the second planet does not 
transit and/or the second planet is small enough to be undetectable (initially) in transit. This latter case might occur 
when the planets are in resonance, since this significantly amplifies the TTV signal. 

There is also some information in the changes in transit durations or shapes due to additional planets, but \citep{2009ApJ...698.1778R} conclude that, in the majority of expected exoplanet systems, the 
signal due to transit timing anomalies is usually stronger than that due to variations in the transit shape, unless 
the planets are in specific non-resonant orbits with large mutual inclinations. We confirm this notion and have 
found, for example, that the 
amplitude of transit duration variations (TDVs) for HAT-P-13b as induced by HAT-P-13c are less than half the size of the 
transit timing anomalies, at any relative inclination. Note also that the accuracy of transit duration measurements is 
usually less than half as good as transit timing measurements \citep{2008ApJ...689..499C} and that transit durations are much more sensitive to possible systematic effects such as observing in different filters. Although ``transit shaping'' is 
harder to detect than transit timing, when it is reliably detected, it can have useful value for determining the 
properties of the planetary system. Note that when the star experiences significant acceleration during the course of a transit, TTVs and TDVs may not be entirely well defined (see 
$\S$\ref{model} below).

Beyond S10, a few studies have considered systems where more than one planet is found to transit. Anomalies in photometry from TrES-1b from Hubble and the ground were conjectured to be possibly due to an additional planet \citep{2009A&A...494..391R}, although the
authors themselves acknowledge this as an unlikely possibility. \citet{2009ApJ...701..756D} argue that such signals are much better explained
by crossing starspots, especially for TrES-1, which is known to be photometrically active. Authors have noted that nearly coplanar multi-planet systems are relatively likely ($\gtrsim$10\%) to have multiple planets transiting \citep{2004ESASP.538..407S,2009IAUS..253..173F,2005Sci...307.1288H}. \citet{2004ESASP.538..407S} also quantitatively showed that confusion between which planet is transiting at any particular time is extremely unlikely.

\citet{2005astro.ph..1440M} submitted a manuscript to arXiv that was never published, with a claim of observing a transit of a second resonant planet in OGLE-TR-111 system; new observations by \citet{2010ApJ...714...13A} have shown that there are no significant TTVs in this system.

Though S10 report five new candidate multi-transiting systems, as far as we are aware, there is no 
known instance of an object that transits or eclipses one or both stars in an eclipsing binary (D. Fabrycky, G. Torres, pers. comm.), though many eclipsing binaries are known to have additional companions (i.e., from the stellar equivalent of 
TTVs). Eclipses of the same star by two stellar companions are \emph{a priori} unlikely since the large masses of three stars imply that gravitational stability is only possible if the triple is hierarchical, requiring the outer star to have a large orbit and making eclipses observable from Earth relatively improbable. Despite early searches \citep[e.g.,][]{1998A&A...338..479D}, to date, no transiting planets have been found around eclipsing binaries, though \citet{2009IAUS..253..382A} propose an algorithm for detecting such systems.

\subsection{Geometric Probability of Observing Multi-Transiting Systems}
\label{probmulti}
From radial velocity observations, there is growing evidence that hot Neptune/Super-Earth planets with short periods ($P \lesssim 50$ days) are common, with a reported frequency around 30\% \citep{2009A&A...493..639M}. It is also apparent that hot Neptunes tend to cluster in multiple systems \citep{2010A&A...512A..48L} and that these could potentially show multiple planets transiting (S10). This indicates that the probability of finding multi-transiting systems, especially using with the precise photometry and long baseline of \emph{Kepler}, will be relatively high. Can we place a quantitative estimate on the number of multi-transiting systems we expect?

While a full answer requires a detailed model of planet formation and evolution \citep[e.g.,][]{2010arXiv1006.2584I}, we can straightforwardly address the geometric aspect of this question. In particular, we consider the following. Suppose that all stars that have a transiting planet (1) also have an additional planet (2) at a true mutual inclination of $\cos \phi_{12} \equiv \cos i_1 \cos i_2 + \sin i_1 \sin i_2 \cos (\Omega_1 - \Omega_2)$. (These angles are defined in $\S$\ref{model}.) What is the probability that, given planet 1 is transiting, that planet 2 will also be transiting, as a function of $\phi$ and the semi-major axes of the two planets?

\begin{figure}						 
\begin{center}
\includegraphics[width=\columnwidth,trim=0in 0in 0in 4.5in]{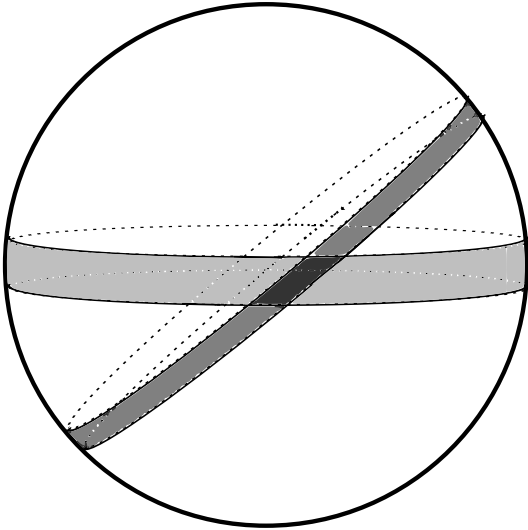}
	\caption{\label{fig:sphere} Illustration for determining frequency of multi-transiting systems. The celestial sphere, centered on the parent star, can be divided into regions where 
no planets are observed to transit (white), only one planet transits (lighter shade of gray), and both planets transit 
(darkest shade of gray). (Only parts of the closer hemisphere of celestial sphere are shaded in this illustration.) In a latitude band centered on the planet's (current) orbital plane, distant observers would 
see the planet 
transit; the size of this band is given by $\frac{R_*}{a}$, in the case of circular orbits, with the more distant 
planet having the narrower band. An additional planet is inclined at the true mutual inclination $\phi$. In the overlap 
region (dark gray), both planets are seen to transit. The apparent on-the-sky inclinations for the planets depend on 
the 
exact location of the observer in this region. The area of this region, divided by the surface area of the sphere, 
gives the probability of seeing both planets transit. See discussion in text and Figure \ref{fig:prob}.
         }
\end{center}
\end{figure}

To answer this geometric question, we turn to the method of \citep{1984Icar...58..121B}, later repeated by other authors, who consider the fraction of the celestial sphere centered on the parent star where distant observers would see the planet(s) transit. A model of this is shown in Figure \ref{fig:sphere}. There is a certain fraction of the celestial sphere where distant observers would see no planets transit, the inner planet transiting, the outer planet transiting, or both planets transiting. This depends both on the semi-major axis (in stellar units) of each planet, as well as the true mutual inclination. There are two cases: 1) the mutual inclination is low, so that the entire region that can see transits by the outer planet can also always see transits by the inner planet and 2) the mutual inclination is high enough that observers must be near the line of nodes (the intersection of the two orbital planes) to observe both planets transiting. 

In the low mutual inclination case, the probability that both planets are seen to transit is just $R_*/a_2$, the transit probability of the outer planet. In the high mutual inclination case, an estimate of the probability can be made by ignoring the spherical geometry and calculating the area of the parallelogram on the celestial sphere near the line of nodes where a distant observer would see both planets transit (Figure \ref{fig:sphere}), which gives
\be
p(1\&2) = \frac{R_*^2}{a_1 a_2 \sin \phi}
\ee
where $p$ is used for probability and ``1\&2'' means both planets transit. 
Both geometrically and from continuity, it can be seen that the  inclination which divides these regimes is $\phi_{\rm crit} \simeq \sin \phi_{\rm crit} \simeq \frac{R_*}{a_1}$, although this is approximate as it also depends somewhat on the spherical geometry of the system. 

\begin{figure}						 
\begin{center}
\includegraphics[width=0.7\columnwidth,angle=90,trim=0in 0in 1in 1in]{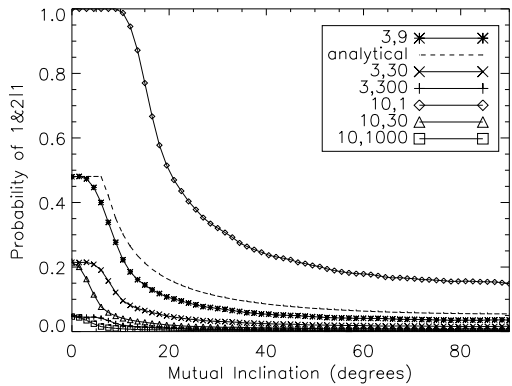}
	\caption{\label{fig:prob} 
Based on the geometric model 
presented in Figure \ref{fig:sphere} and described in the text, 
the probability that a two-planet system would show two planets transiting, assuming that the first planet transits, 
can be calculated as a function of the true mutual inclination $\phi$ (which is only weakly related to the apparent 
inclination measured by transit durations) ranging from coplanar ($0^{\circ}$) to perpendicular ($90^{\circ}$); the 
curves are symmetric around $\phi=90^{\circ}$. It is important to note that the calculated probability assumes that 
all systems are two-planet systems, i.e., is a geometric 
argument only and does not say anything directly about the astrophysical likelihood of double-transiting systems. The 
solid 
lines give the probability (denoted 1\&2|1) for various systems as determined by a Monte Carlo model, which accounts 
for a random viewing angle. The 
planetary systems in question are identified by the legend: the first number 
is the orbital period in days of the planet that is assumed to transit and the second number is the orbital period in 
days 
of the planet whose transit probability is being assessed. For example, a system with a 3-day transiting planet and an 
additional planet at 9 days is shown with asterisks. An analytical estimate, divided into two inclination regimes, is 
described in the text and given by the dashed line for the system with planets at 3 and 9 days. We have also shown the probability that a 1-day planet transits assuming that a 10-day outer 
planet is known to transit (diamonds); at low mutual inclinations, transits of the 1-day planet are assured. Note that 
the reciprocal of the probability given above is the number of similar systems that would be expected in the full 
population where the second is present but not seen to transit.
         }\end{center}
\end{figure}

These are the probabilities of observing both planets in transit given no a priori information. If we know that one planet is transiting, then we can calculate the probability of the other planet transiting given that one planet is known to transit. That is, we can now answer the proposed question, what is the probability that the planet 2 will transit if planet 1 is known to transit. The approximate answer is (see also Figure \ref{fig:prob}): 

\[p(1\&2|1) \simeq \left\{
\begin{array}{l l}
1 & \mbox{if $a_1 > a_2$ and $\phi \lesssim \frac{R_*}{a_1}$} \\
\frac{a_1}{a_2} & \mbox{ if $a_1 < a_1$ and $\phi \lesssim \frac{R_*}{a_1}$} \\
\frac{R_*}{a_2 \sin \phi} & \mbox{if $\phi \gtrsim \frac{R_*}{a_1}$}
\end{array} \right. \]

These analytic expressions are shown with the results of a simple Monte Carlo model in Figure \ref{fig:prob}. In this model two 
planets are set up with mutual angles $\phi$, and are observed from $10^5$ lines of sight placed randomly on the celestial 
sphere. The model then calculates the apparent on-the-sky inclinations, and then gives the probability that both planets 
are seen to transit if the planet 1 is seen to transit. We ignore the size of the planet and non-circular orbits; these can be incorporated by adjusting the size of the star or, for eccentric orbits, using the ``effective'' semi-major axis that gives the same star-planet distance $r=a\frac{1-e^2}{1+e \sin \omega}$ \citep{2010arXiv1001.2010W}. We also assume the planets to be on fixed orbits, though interactions over many orbital periods can result in precessing orbits that go in and out of transit \citep[e.g.,][]{1995EM&P...71..153S,2006ApJ...653..700S}. The two regimes can be seen and the overplotted analytical expressions provide a reasonable approximation; differences between the analytical and numerical calculations result from not strictly incorporating the spherical geometry and diminish with increasing semi-major axes. Our result is comparable to, but more general than, the derivation of the double-transiting probability by \citet{1996chz..conf..229K} and \citet{2005Sci...307.1288H} because we include the appropriate random distribution of the inclination of the transiting planet. These analytical expressions suggest that the probability of seeing an both Venus and Earth transit, given an exact analogue, is $\sim$0.06\%, though if the Venus analogue is known to transit, then the probability of the Earth analogue also transiting is 8.4\%. For the Super-Earth system triple HD 40307, if the mutual inclination is low ($\lesssim$1.5$^{\circ}$), then the probability that all three transit is essentially just the probability that the outer planet transits, $\sim$2.6\%. The probability of observing systems like those recently announced is given in S10.

We have verified that the above results are consistent with the random 3-dimensional placement of the Laplace plane and a random orientation of the two planets with respect to that plane. Assuming a two planet system, the inclination distribution (from the plane of the sky) of the second planet can be determined by:
\be
\cos i_2 = \cos i_1 \cos \phi + \sin i_1 \sin \phi \cos \Delta
\ee
where $\Delta$ is a randomly chosen angle (and inclinations are in the observer's reference frame, $i = 90^{\circ}$ is 
edge-on). The distribution of $i_2 - i_1$ is not Gaussian. Indeed, it is peaked at the extremes: $i_2$ is most likely to be $i_1 + \phi$ or $i_1 - \phi$, though there is still significant probability throughout the range. (Inclinations, $i_2$, greater than 90 degrees are equivalent to $180^{\circ}-i_2$.) Note that this accounts for the full three-dimensionality of the problem, rather than assuming a typical on-the-sky inclination difference (which is only weakly related to the true mutual inclination) with some distribution \citep{2008arXiv0804.1150B}. The above discussion assumes there is no systematic relationship between the orbits of the planets, but this is not necessarily true when the mutual inclination is in the Kozai regime ($40^{\circ} \lesssim \phi \lesssim 140^{\circ}$), where the secular Kozai effect can affect the orientation of the planetary orbits \citep{1962AJ.....67..591K,1962P&SS....9..719L}. In particular, the Kozai effect acts to bring the periapse of inner planet to occur when it is maximally out of the plane of the outer planet; this would effect the probability of observing this system as double-transiting and should be kept in mind. 

Based on these results we can draw the following inferences. The fraction of double-transiting systems remains relatively high ($\frac{a_2}{a_1}$) even for non-zero mutual inclinations ($\phi \lesssim \frac{R_*}{a_1}$). The true number of multiple systems with one planet transiting is usually a few to several times larger than the number of systems where both planets are seen to transit, even if all systems are nearly coplanar. This reinforces the fact that the search for TTVs is still an important method for probing multiple planetary systems. In particular, the types of systems where multiple planets are seen to transit will be strongly biased against those with large ($\gtrsim5^\circ$) mutual inclinations. Even if the fraction of double to single systems detected by \emph{Kepler} is comparable to the values expected from nearly coplanar orbits, this does not necessarily imply that all systems have low mutual inclinations. One should compare the distribution of other orbital parameters (e.g. eccentricities, spin-orbit alignments) of multi-transiting systems with, for example, radial velocity multiple systems as geometric biases imply that multi-transiting systems are drawn from the low-inclination subset of multiple planet systems. Statistical techniques, such as those employed in \citet{2009ApJ...696.1230F}, can make more rigorous statements about the distribution of mutual inclinations with strong implications for planet formation as discussed in $\S$\ref{mutinc}.

In systems with more than two planets, secular evolution of the mutual inclination is possible and $n$-body interactions (i.e., as characterized by second-order Laplace-Lagrange theory) are relevant to the evolution of the planetary inclinations and nodal longitudes. This implies that the orbital plane may be constrained in ways not considered here.

\section{The Value of Multi-Transiting Systems}
\label{valuemts}
Multi-transiting systems are valuable for understanding the physical and orbital parameters of distant planetary systems for at least two reasons. First, since both planets transit, we have a wealth of information about each planet that is not known for non-transiting planets, such as the radius and inclination, or is much more precisely determined, such as orbital period and phase. This already makes multi-transiting systems quite valuable. The second valuable aspect is that both planets are known to transit the same star and these planets are often close enough to affect one another's orbit. This synergy enhances the value of these systems and the following discussion attempts to focus on this aspect of multi-transiting systems. In particular, we consider the value of such systems in eliminating false positives, measuring relative and absolute masses and radii, determining mutual inclinations, providing additional dynamical insights, and for use in comparative planetology.

\subsection{Eliminating False Positives}
\label{falsepos}
Many transiting systems, especially those discovered by \emph{Kepler}, will be around faint, active, and/or rapidly rotating stars for which it is 
difficult to get abundant and accurate radial velocities. Observation time to confirm the hundreds of good planetary candidates from \emph{Kepler} \citep{2010arXiv1006.2799B} will also be very limited, esepcially given the observed preference for lower mass planets. In the case of single transiting planets, radial velocity 
confirmation is generally used to show that the light curve is not created by any other astronomical phenomenon and to measure the minimum mass of the putative companion.
Astrophysical signals that mimic the light curve of transiting planets (i.e., producing a $\sim$1\% drop in light on 
several hour timescales) have been a serious issue for ground-based 
transit surveys. These astrophysical ``false 
positives'' are usually very common and include, for example, an eclipsing binary contaminated by the third light of a 
brighter (possibly related) star, a dwarf star orbiting a giant, or a grazing eclipsing binary\footnote{We will not 
discuss here false positives due to instrumental or systematic effects, which also need to be carefully ruled out.}. 
With its astounding 
photometric and astrometric precision, \emph{Kepler} can eliminate many, but not all, of these false positives even before 
spectroscopic investigation \citep{2010ApJ...713L.103B}. In any case, the hypothesis that the light curve is caused by transiting planets 
can either be verified through radial velocity measurements that measure the mass or by eliminating all plausible 
astrophysical false positives. This section will be concerned with the latter technique, since many transiting planets 
will not be able to be fully characterized with radial velocities due to astrophysical or logistical reasons.

Multi-transiting systems are more difficult to mimic astrophysically than single-transiting systems, though it is not impossible. Some examples 
include, two eclipsing binary systems, either both grazing or contaminated by each other's light (or the ``fifth'' light 
of another star). Such systems are rare, but not unheard of \citep{2007IBVS.5768....1P}. Multi-transiting systems can 
also be comprised of a combination of false positives, planets around different stars, and planets around the same star. For this reason, each candidate transit signal should 
be investigated separately. Note that there are two steps in eliminating 
false positives in multi-transiting systems: showing that the candidates transit the same star and then investigating 
the false positives in the case of multiple objects orbiting the same star.

With precise photometry and astrometry, as is available from \emph{Kepler}, several signs of false positives can be 
investigated: a secondary 
eclipse that is too deep to be caused by a planet; ellipsoidal oscillations indicating high 
masses; ``Doppler photometry'' or relativistic beaming from the stellar motion, that indicates high masses \citep{2003ApJ...588L.117L}; multiple distinct stellar rotation or pulsation periods; variability typical of giant stars \citep{2010ApJ...713L.120J}; astrometric centroid motion during eclipse, indicating a 
background eclipsing binary \citep{2010ApJ...713L.103B}; and other methods. Most of these techniques appear to be used 
regularly by the \emph{Kepler} team (S10) and should be applied to each individual planet candidate with the requirement of consistent results, e.g., apparent astrometric motion due to crowding should go in the same direction for both candidates. We note that \emph{Kepler} here does have one disadvantage compared to CoRoT in that it measures only mono-chromatic light curves and observations in multiple colors 
are an excellent way of ruling out false positives \citep{1971Icar...14...71R,2009A&A...506..287L}. Observing even one transit of each candidate in a different filter is an excellent way to rule out false positives, besides the added benefit of helping to break degeneracies involving limb darkening \citep{2009ApJ...703.1086C}. Observations of the Rossiter-McLaughlin effect can also be used to rule out false positives.

There are a few techniques for eliminating false positives that are possible only in multi-transiting systems. To show 
that the two planet candidates orbit the same star, one can check that the stellar properties inferred from the 
transits are consistent between each other and with the expectations for the stellar properties (e.g., based on 
spectral type). For example, the durations of the transits compared with the orbital periods can be used to 
estimate the stellar density \citep{2003ApJ...585.1038S}. This technique is not as powerful as might be expected 
because, though the stellar density can be measured precisely by a transiting planet on a circular orbit (if the impact parameter is well determined), 
allowing for a non-zero eccentricity introduces significant uncertainty in the stellar density measurement, which, to 
first order in eccentricity, is proportional to $(1+3e)$ (S10). There is a maximum eccentricity for each planet, assuming 
they are in orbit around the same star, from stability constraints. Therefore, if the inferred stellar densities from 
each planet are more different than allowed given the eccentricity constraints, it can be assumed that the candidates 
do not orbit the same star. Similarly, inferred limb darkening coefficients, if known accurately and without 
degeneracy, must be similar if the candidate planets are transiting the same star. While these are important ways of 
identifying false positives, particularly for high signal-to-noise planets, in practice, these properties alone do not seem sufficient for proving that candidates are indeed planets orbiting the same star.

A much stronger indication that the candidates orbit the same star is the observation of TTVs, TDVs, or mutual events ($\S$\ref{mutev}). This is true 
for single-transiting systems as well, but in multi-transiting systems if the TTVs are strongly consistent with 
perturbations from the other known transiting object, then this is a strong indication that both candidates orbit the 
same star. Care needs to be taken here to note that TTV measurements can be quite degenerate and/or affected by additional unseen perturbing planets. Interdependent TTVs that are consistent with each object perturbing one another provide clear evidence that 
both objects are in orbit around the same star.  Whether TTVs are 
observed for one planet or more, if the masses of the 
planets may also be measured or constrained, this is equivalent to 
radial velocity mass measurements and proves the 
planetary hypothesis (see $\S$\ref{dynmass} below). Indeed, once it is known, through whatever means, that both candidates 
orbit the same star, then a measurement or upper limits on TTVs will usually be sufficient to rule out the possibility 
that one or more of the candidates is stellar in nature. Orbital stability can also be invoked and will often be enough 
to rule out stellar masses. This is because stars are $\sim$10$^{2-4}$ times more massive than planets in general; brown dwarfs will
be more difficult to rule out. Orbital stability can be used to identify false positives as well: if 
two candidates have very similar periods, then their orbits may be too close to one another to be stable, even with 
low planetary masses. These can be a chance superposition. 

It is important to note here that numerical analyses of multi-transiting systems, i.e. using the model described in 
$\S$\ref{model}, are a valid and valuable method of ``follow-up'' to confirm the true planetary nature of these systems.

Continuing along this theme, there are some expectations for the orbital spacing in multi-planet systems both 
observationally from the known RV planets and theoretically from planet formation simulations (though if the field of 
exoplanets has taught us anything, it is that planet formation theories may be incomplete). For example, planets in the same system are much more likely to lie near mean motion resonances through various formation and evolution effects, while planets around different stars will only appear to be nearly resonant by coincidence \citep[compare to][]{1965MNRAS.130..159G}. Near-resonant architectures are already seen in the candidate multi-transiting systems of S10 and are a good indication that these systems are planetary in nature. System architecture is especially 
convincing in multi-transiting systems with more than two planets. Imagine a quadruple-transiting system that passes 
all the basic tests for false positives for each of the candidate planets. Any non-planetary scenario
contrived to match the light curve will have ridiculously low probabilities: it is not reasonable to consider 8 equal 
brightness stars each in perfect eclipsing binaries with randomly selected periods having typical planetary spacing. You can then 
say with confidence that at least some of these candidates must be around the same star. Similarly, systems with apparent orbital periods that are so close together that stability is difficult may be an indication of a chance superposition of candidates around different stars. Also, given our understanding on the mass 
dependence of multiple systems as discussed above, a system with multiple Neptune-size planets seems more likely than a system
with multiple Jupiter-like planets.

Even after the planetary nature of a candidate is confirmed, blending can be an issue for interpretation. For example, a transiting Jupiter blended with a brighter star will have the depth of a much smaller planet. However, the full transit shape of a blended Jupiter and a true Neptune are not exactly the same (e.g., different ingress/egress durations) and \emph{Kepler} is often powerful enough to resolve this degeneracy to some extent. 

Finally, we should note that systems which contain a false positive can still be scientifically interesting (e.g., a 
planet in an eclipsing binary).

\subsection{Dynamical Masses and Absolute Radii}
\label{dynmass}
Kepler's third law relates the measured semi-major axis and period to the sum of the masses, but does not provide any information on each mass individually. In gravitationally interacting three-body systems, the degeneracies between the masses are broken and each of the mass ratios between the objects can be obtained independently. This is valuable for several reasons. 

First, small mass ratios imply that the companions must be planetary in nature. That is, measuring mass ratios from TTVs is an extremely valuable method for rejecting false positives, especially when interdependent TTVs are simultaneously measured, as discussed above. 

Second, \citet{2005MNRAS.359..567A} point out that adding radial velocities to the transit timing variations gives dynamical masses for all of the objects in the system. Photometry alone combined with Kepler's third law measures the system parameters well, but the solution is invariant under transformations that conserve the stellar density, mass ratios, and radius ratios. In single-transiting systems, the mass ratio between the planet and the star must be provided by stellar RV measurements and the stellar mass is inferred through stellar modeling; often the the precision in absolute planetary parameters is limited by uncertainties in the stellar models. However, when TTVs accurately measure the mass ratio, the RV signal can be directly converted in the dynamical mass of the star \citep{2005MNRAS.359..567A}. From the photometry, the stellar density can be precisely measured \citep{2003ApJ...585.1038S}, so the radius of the star can also be determined in an absolute sense. Combining this with the precise radius ratios leads to very well determined radii for each object. When TTVs and RV are used in concert, there is no need for stellar models; indeed, such systems could be used as standards to improve stellar and planetary models. That the combination of TTVs and RVs yields dynamical masses was confirmed numerically with simulated data by \citet{2010arXiv1005.5396M}.

It is also possible to solve for dynamical masses without using radial velocities at all. If the light-travel timing offset has a detectable TTV signal, it sets an absolute scale and measures dynamical masses without RVs. Generally, these shifts will be small unless the planets are massive, eccentric, and/or have strong TTV signals (Agol et al. 2005; Veras \& Ford 2010, submitted). Single-transiting planets have a small light-travel time effect during the course of the transit \citep{2005ApJ...623L..45L} that is not detectable in practice even with \emph{Kepler} photometry \citep{2009ApJ...698.1778R}. However, multi-transiting systems provide both a precise clock (the inner transiting planet) and a well-known perturber (the outer transiting planet), so these are the systems that are most likely to have dynamical masses measurable from light-travel time TTVs; in these systems, dynamical masses are measured by photometry alone. Note that the light-travel time effect has to be measured with high accuracy in order to provide a useful limit on the overall scale. It is also important to keep in mind the effects of potential unseen planets, particularly distant planets, which can confuse the interpretation of the light-travel time offset. Each system will need to be considered independently to see whether RV observations can easily provide a tighter constraint. 

The stellar radial velocity also has a small photometric effect due to relativistic beaming (at quite non-relativistic speeds), which is known as photometric Doppler boosting \citep{2003ApJ...588L.117L}. The photometric amplitude of this signal is usually a few times the radial velocity signal divided by the speed of light, e.g., a large RV signal for a very hot Jupiter of $\sim$1 km/s would result in a photometric amplitude of $\sim$10 parts per million (ppm). On a faint \emph{Kepler} star ($V \approx 14$,$\sigma \approx 1000$ ppm/min), the photometric variation of such a best case system could be detected at the 3-$\sigma$ level in 21 days, but only in the absence of astrophysical noise. With planets that are smaller, longer period, and in the presence of astrophysical noise sources with comparable amplitudes (spotted stellar rotation, reflected light curves, ellipsoidal oscillations, etc.), this detection will be much more difficult. Even so, it is worth noting that the combination of this effect with TTVs again could, in theory, provide a measurement of dynamical masses and absolute radii from photometry alone. 

\subsection{Mutual Inclination}
\label{mutinc}
The extent to which
extra-solar planets are mutually inclined is a strong constraint on theories of planet formation and in understanding the evolution of
planetary systems, whether by disk-induced migration \citep[e.g.,][]{2007ApJ...654.1110T}, planet-planet scattering \citep{chat2008,juri2008,2010ApJ...714..194M}, Kozai effects \citep[e.g.,][]{2007ApJ...669.1298F}, or
other mechanisms. Each of these will produce a different inclination distribution, and, when coupled with the eccentricity distribution
and measurement of the Rossiter-McLaughlin effect ($\S$\ref{multiRM}), promises the answer to many of our questions about the origins of planetary systems \citep{2009IAUS..253..173F}. It is important to note here that observations of multi-transiting systems do not immediately reveal the actual mutual
inclination of the planets, which is the quantity of greatest interest \citep{2009IAUS..253..173F}. Even planets that have large mutual
inclinations can lead to multi-transiting systems in the correct orientation, though the probability of observing systems shrinks
rapidly with large mutual inclinations ($\S$\ref{probmulti}). This implies that the number and distribution of multi=transiting systems, as an ensemble, will be extremely valuable for increasing our understanding of the formation and evolution of planetary systems (S10). The importance of multi-transiting systems in constraining the mutual inclinations was recognized early, \citep[e.g.,][]{1971Icar...14...71R,1984Icar...58..121B,1996chz..conf..229K}

The astrophysically valuable angle is the true mutual inclination $\cos \phi_{12} = \cos i_1 \cos i_2 + \sin i_1 \sin i_2 \cos (\Omega_1-\Omega_2)$ is a powerful indicator of past evolution, but it is not measured directly even in multi-transiting systems (see $\S$\ref{model} for definition of angles). It can only be weakly constrained as $i_1 + i_2 \le \phi \le 180^{\circ} - i_1 - i_2$, with most of the likelihood at the extremes of this distribution, but significant possibility anywhere in that range. However, there are several techniques that can improve our knowledge of the actual mutual inclination. 

One way of measuring $\phi$ is to resolve the orbit on the sky through astrometry \citep[e.g.,][]{2010ApJ...715.1203M},
interferometry \citep{2008PASP..120...38U}, polarimetry \citep[e.g.,][]{2009ApJ...696.1116W}, or direct imaging \citep{2008Sci...322.1348M}. This can measure the nodal angles directly and, in
combination with photometry, could yield good constraints on the mutual inclination. With current observational technology, these
techniques are only possible in nearby and/or bright systems, and measurement of precise mutual inclinations is rare, though future instruments may make these observations possible. The mutual inclination for the resonant  
planets in GJ876 was measured from an extensive and long set of radial velocity measurements \citep{2010A&A...511A..21C,2009A&A...496..249B}, but this method is not generally applicable to other systems and, although the true masses of the planets are now known, their radii and densities are not known since they are not transiting.

Generally a measurement of absolute orientation on the sky is not possible, and so a more ``indirect'' measure of the mutual inclination will be required. For multi-transiting systems, since $\phi$ only depends on the mutual nodal angle, any measurement that can determine the relative nodal position ($\Omega_1 - \Omega_2$) also measures the true mutual inclination; we will make use of this fact in our discussion below. It is already known that mutually inclined systems undergo nodal precession which
changes the observed impact parameters of the systems and significantly changes transit shapes \citep{2002ApJ...564.1019M}. 
Mutual inclinations also can be constrained by TTVs \citep{2009ApJ...707..446B,2009ApJ...701.1116N,2010ApJ...712L..86P} which is most readily applied in multi-transiting systems. With the knowledge of planetary
$M \sin i$ from radial velocities, the lack of TTVs or TDVs and/or a requirement on orbital stability can give an upper limit on the mutual inclination \citep{2004ESASP.538..407S}. We present new promising methods of measuring mutual inclinations of multi-transiting systems in $\S$\ref{othersec}.

\subsection{Comparative Planetology}
\label{comppl}
Another value of multi-transiting systems was pointed out by \citet{2005astro.ph..1440M}: uncertainties in the stellar mass and radius
propagate equally to all the planets, meaning that the inferred mass and radii of the planets (whose uncertainties can be dominated
by uncertainties in the stellar properties) can be compared more directly. It is also exciting to note that planets in the same
system are likely to have some comparable properties (e.g., age and perhaps metallicity), so that comparison of internal models for planets in the same system could be a
more powerful discriminant of interior processes than comparisons between planets around different stars \citet[e.g.,][]{2010EAS....41..355G}. The effects of irradiation or tidal heating can be seen more strongly in
contrast since many of the other variables are probably consistent from planet to planet in the same system. Furthermore, correlations between the planetary system architecture (where planets are relative to one another) and the bulk composition of these planets as inferred from their densities will also be quite revealing, especially in systems with three or more planets (S10). In our own solar system, the rocky planets are interior to the gas giants which are interior to the ice giants, but exoplanetary systems (that have likely undergone migration to increase their transit probability) may show interesting differences. 

From geometric constraints ($\S$\ref{probmulti}), multi-transiting systems are likely to have
relatively small orbital periods, so that tidal interactions with the parent star could be non-negligible. Detailed orbital knowledge
derivable from multi-transiting systems will help us understand the interplay between 
tidal
forces and secular planetary perturbations and potentially measure the Love number $k_{2p}$ (which probes the planetary interior structure) and tidal dissipation parameter $Q_p$ for the inner planet \citep{2002ApJ...564.1024W,2007MNRAS.382.1768M,2009ApJ...704L..49B,2010arXiv1001.4079M}.

The Love number is a unique and novel probe of the planetary interior that provides additional information beyond the planetary density as a measure of central condensation \citep{2009ApJ...698.1778R,2009ApJ...707.1000H}. The aforementioned method for measuring $k_{2p}$ is indirect and only works for small inclinations \citep{2010arXiv1001.4079M} and with planets whose orbits are very well characterized; multi-transiting systems are the best opportunity for these inferences.

The tidal love number $k_{2p}$ can also be measured directly by observing apsidal motion, a well-known practice in the stellar binary community that has recently been applied to exoplanetary systems \citep{2009ApJ...698.1778R}. Since multi-transiting systems may be well characterized with 
dynamical masses not resulting from the direct application of Kepler's Law, it is 
plausible that some multi-transiting systems can be used to 
constrain or measure additional non-Keplerian parameters\footnote{The best multi-transiting systems may also potentially serve as tests of 
modified theories of gravity \citep[e.g.,][]{2009arXiv0909.5355I}. 
} including general relativity or $k_{2p}$ for the interior planet \citep{2009ApJ...698.1778R}. The unique determination of limb darkening by multiple planets in transit may also improve the possibility of measuring planetary asphericity, another probe of $k_{2p}$, which is usually degenerate with limb darkening parameters in single-transiting systems \citep{2003ApJ...588..545B,2009ApJ...698.1778R,2010ApJ...709.1219C}. In all, we expect that multi-transiting systems will be very useful for studying diverse and unknown exoplanetary interior structure.

We also note that, in systems where one planet is already known to transit, single events from very long period planets ($P \gtrsim 1$ year) can be interpreted accurately. These same events around random stars are less robust and less likely, thus multi-transiting systems provide a unique opportunity to probe the (very) rare distant planets that transit their parent star.

\subsection{Additional Dynamics}
\label{adddyn}
In systems where more than one transiting planet is known to be orbiting the same star, the durations of central transits by planets with circular orbits have a fixed ratio. Indeed, it is this fact that allows for a precise determination of the stellar density from transit light curve modeling \citep{2003ApJ...585.1038S}. While allowing for eccentric orbits and non-central transits makes this more difficult to use as a technique for identifying false positives (S10), once multiple planets are known to be orbiting the same star, the transit durations can be used the other way as a constraint on
eccentricities. This is because the period ratio is well known, giving
the semi-major axis ratio and hence the orbital velocity ratio. As this
velocity is measured by transit duration, as long as the transits have
high signal-to-noise, deviations from the expected ratio can be attributed
to faster or slower motion due to the eccentricities of the bodies.

From Eq 29 in \citep{2010arXiv1001.2010W}, we find that the well determined ratio is:
\be
\left( \frac{1 + e_b \sin \omega_b}{1 + e_c \sin \omega_c} \right) \left( \frac{1-e_c^2}{1-e_b^2} \right) = \left( \frac{P_b}{P_c} \right)^{5/3}
 \left( \frac{\delta_b}{\delta_c} \right)^{1/4} \left( \frac{T_c \tau_c}{T_b \tau_b} \right)^{1/2}
\ee
where the transit parameters that are well determined are the periods ($P$), depths ($\delta$), duration ($T$), and often ingress/egress times ($\tau$). This ratio is often very well determined assuming there is no degeneracy between limb darkening coefficients and impact parameters (which is not necessarily true for \emph{Kepler} planet); in the case of small
eccentricities we expect a good constraint on $e_b \sin \omega_b - e_c
\sin \omega_c$. If this value is large, then it is likely that one of the
planets is eccentric. If $\rho_*$ is well known from asteroseismology \citep{2010ApJ...713L.169C,2010ApJ...713L.164C}, then $e \sin \omega$ can be inferred for each transiting planet (in single or multi-transiting systems).
TTVs are also quite sensitive to eccentricities and orientations \citep{2009ApJ...701.1116N,2010ApJ...712L..86P}

Furthermore, if there are multiple large close-in planets, then it is reasonable to assume that these planets must have migrated inward, perhaps
together. Theories of multiple-planet migration would predict that these are probably in or near resonances \citep[e.g.,][]{2007ApJ...654.1110T}, and TTVs are much more sensitive to near-resonant planets than radial velocities \citep{2005MNRAS.359..567A,2005Sci...307.1288H}. Multi-transiting systems will be the best for understanding these TTV signals without degeneracies, which are present even in well-characterized resonant systems \citep{2010arXiv1005.5396M}. The presence of additional planets with known periods also limits eccentricities (past and present) due to stability constraints.

\section{A Full Numerical-Photometric Model}
\label{model}
To better understand the value of multi-transiting systems, we have developed a model that combines an $n$-body integration with calculation of the photometric diminution from each of the planets. This full numerical-photometric model can be applied directly in the analysis of light curves with interacting planets, including multi-transiting systems like those discovered by S10. While no such analysis is performed in this work, we will comment on the value of this model in fitting actual data. 

We use the coordinate system suggested by Fabrycky 2009. The orbits are described using astrocentric equinoctial 
elements, which avoid the singularities of Keplerian elements. The equinoctial elements are \citep{1970AIAAJ...8....4A} 
$a$, $h = e \sin \varpi$, $k = e \cos \varpi$, $p = \tan (i/2) \sin \Omega$, $q = \tan (i/2) \cos \Omega$, and 
$\lambda^{ml} = M + \varpi$, where $a$ is the semi-major axis (the same in Keplerian and equinoctial 
elements), $e$ is the eccentricity, $\varpi \equiv \omega + \Omega$ is the longitude of 
pericenter, i.e. the sum of the argument of pericenter ($\omega$) and the longitude of the ascending node ($\Omega$), 
$i$ is the inclination (with respect to the plane of the sky), and $\lambda^{ml}$ is the mean longitude, i.e. the sum 
of the mean anomaly ($M$) and the longitude of pericenter. Note that $h$ and $k$ are sometimes defined in the opposite 
way and that $\lambda^{ml}$ is not related to the angle $\lambda$ associated with the Rossiter-McLaughlin effect (discussed in $\S$\ref{multiRM}).

The longitude of the ascending node, $\Omega$, in binary stellar systems is defined as the position angle of the point where the companion pierces
the plane of the sky coming toward the observer and is measured with respect to astronomical North \citep{1973bmss.book.....B}. This angle
can only be determined if the system is somehow resolved on the sky, e.g., via astrometry, interferometry, polarimetry, or direct imaging. Generally
speaking, this angle will not be known, and without loss of generality we set $\Omega$ of the innermost planet to be equal to
0$^{\circ}$. The difference between the nodal longitudes of the planets \emph{is} detectable, so $\Omega$ for the remaining planets
remains a fitted parameter.

It is possible that the innermost planets are bright enough to be detected in secondary eclipse or in the full orbital phase curve. For
simplicity, we ignore the phase curve since it has more free parameters, but in practice it is not difficult to include in a parametrized sense \citep{2010arXiv1004.3538K}. We also do 
not account for the ellipsoidal oscillations of the parent star, which may be detectable and can also give constraints on the mass 
ratios \citep[e.g.,][]{2010ApJ...713L.145W}. We choose to
keep the possibility of fitting the secondary eclipse, since these eclipses have significant power to constrain the orbital
configuration, most notably the eccentricity \citep{2009ApJ...698.1778R}. Of course, they also contain some information about the atmospheric reflection and
thermal properties of the planet, but this is not our focus. Thus, our model includes a parameter $d^{sec}$ which defines the depth of
the secondary eclipse; if optical and infrared secondaries are both observable, these should each have their own $d^{sec}$ parameters. We note that with current techniques, secondary eclipses can only be detected for planets with periods of $\lesssim$10 days; planets with larger orbital periods have such weak flux that even \emph{Kepler} or \emph{Spitzer} photometry cannot detect them.

The total set of parameters needed to generate photometry of a multi-transiting system of two planets is
$GM_*,R_*,c_1,c_2,c_3,c_4$, $a_1/R_*,h_1,k_1,p_1,q_1,\lambda_1,M_1/M_*,R_1/R_*,d^{sec}_1$, $a_2/R_*,h_2,k_2,p_2,q_2,\lambda_2,M_2/M_*,R_2/R_*$, and, $d^{sec}_2$,
where all the orbital elements are the osculating astrocentric equinoctial elements at a specified epoch (ideally in the middle of the dataset)
and where generally $\Omega_1$ is assumed to be fixed at 0$^{\circ}$ so that $p_1 = \tan (i_1/2) \sin \Omega_1 = 0$ is held fixed. The $c_i$ values are
limb-darkening coefficients, following the notation of \citet{2002ApJ...580L.171M}, which may not all be used depending on the quality of the data. Each additional planet adds 6 orbital parameters, 2 physical parameters (mass and radius ratio), and a parameter for the secondary eclipse depth.

One of the first steps of any data analysis is to first get a good guess of the initial values of all the relevant
parameters. In this sense, multi-transiting systems have the nice property that fitting each planet individually to the data gives
excellent first estimates of nearly all of the parameters described above, especially if photometry and radial velocities are both
used; this is just the normal process used to determine the characteristics of individual transiting planets. Without radial velocities,
there is no constraint on the planetary masses, but a reasonable range of densities can be employed based on theoretical mass-radius relations \citep[e.g.,][]{2007ApJ...659.1661F}. The main parameter that is not well described in this first step is $\Omega_2$, the mutual nodal angle (as $\Omega=0^{\circ}$), which is unconstrained without including the interactions between the two planets. Recall that the mutual nodal angle must be determined to calculate the true mutual inclination between the planets, 
$\cos \phi_{12} \equiv \cos i_1 \cos i_2 + \sin i_1 \sin i_2 \cos (\Omega_1 - \Omega_2)$.

Combining the data from both planets to get a global fit can be done in two different steps. First, the full photometric light curve can be
analyzed returning metadata like central transit times and transit durations, which are then fitted with a fast few-body integrator to refine the relevant parameters, particularly the best-fit masses and $\Delta \Omega$. Transit times are most easily determined using a numerical integration of the
three-body problem, though non-resonant systems can be analyzed much faster using the perturbation method of \citet{2010ApJ...709L..44N}. After estimates for these parameters are narrowed down, the more computationally costly full numerical-photometric model can fit the data directly. Even
when data products like transit times are fit first, the full potential of the data can only be realized with a numerical model, and it
should ideally be the final step in characterizing a multi-transiting system. The drawback to the full model is that it is much more
computationally time-consuming, especially considering the need to explore a highly non-linear $\sim$24-dimensional parameter space. 

Even so, a previously unconsidered aspect of multi-transiting systems may require a full numerical-photometric model to correctly retrieve
the desired parameters in certain systems. Recall that transit timing variations are the result of the aperiodic motion of the relative star-planet position in the
orbital plane. When the relative star-planet position moves out of the orbit plane, transit duration variations results from the change in impact parameter. Similarly, the relative star-planet velocity can change, also causing changes to the shape
of the transit, though this effect is generally weaker. Each of these photometric deviations are approximations to the actual case by
assuming that the star experiences no acceleration during the transit. In certain transiting systems, this may not be a valid
assumption. An extreme example is given by the HAT-P-13 system. The expected duration of the putative transit of HAT-P-13c is 1.28 days
\citep{2009ApJ...707..446B}. This is 44\% of the orbital period of planet b, which is causing the star to executive nearly half of its small
orbit around the star-b barycenter during the transit of c. This affects the relative position of the star and planet $c$ at the 1\% level, which is creates a potentially detectable signature in \emph{Kepler}-quality data. Clearly, the approximation that there is no acceleration during the transit is lost and in such a case, the transit timing
and shaping anomalies are not entirely well-defined. Only a full numerical integration of the motion, coupled directly to the photometric model,
will be capable of a precise description of the data. We note that this new aspect of multi-transiting systems gives some additional
constraint, albeit weak, on the orbit of one planet from its induced stellar motion during the transit of other planet that may not be entirely captured by a TTV/TDV analysis. 

To account for these and other small effects (such as the slightly asymmetric nature of eccentric transits and light-travel time 
effects), a full numerical-photometric model is employed. The model can also easily calculate stellar radial 
velocities and eventually the Rossiter-McLaughlin effect. In our implementation of such a model, the equinoctial 
elements are 
converted to Cartesian coordinates and 
passed to a Fortran 90 $n$-body integrator which integrates the few-body system. This integrator can also be readily 
expanded to include non-Keplerian effects from General Relativity, and the rotational and tidal bulges of the star and 
planet \citep[following][]{2002ApJ...573..829M}, though these are not included in the present analysis.

There can be a difference between the positions of the star and planets in the timeframe of the system and their positions as observed
by the \emph{Kepler} satellite, due to variations in the light travel time. First, we assume that the motion of the Earth or telescope has
been correctly accounted for, i.e. all measurements are referred to the barycenter of the solar system (BJD). Any remaining offset is
due to the motion of the star around the barycenter of the system and, perhaps, motion of the system barycenter relative to the solar
system. This latter effect can have an observable consequence in certain cases \citep{2007ApJ...661.1218S,2009ApJ...700..965R}, but is not considered here.

We found that accounting for the light-travel time effect \citep{2005ApJ...623L..45L} was best accomplished by correcting the positions of the star and each planet to the plane of the sky at the system barycenter at a consistent time. These positions are then used to calculate photometry.

Instantaneous stellar velocities (for comparison to radial velocity data) can be calculated from the positions. For all planets, $i$, all the times at which the on-the-sky distance $d_i=\sqrt{x_i^2+y_i^2}$ is less than the separation needed for transits ($R_i+R_*$), the photometric dimming due to the transit is determined from the codes of \citet{2002ApJ...580L.171M}, using an appropriate limb darkening law and coefficients. 

The quantity of \emph{Kepler} photometry is in a new regime of data fitting that suggests a slightly alternative method for calculating photometric light curves. The codes of \citet{2002ApJ...580L.171M} are fast, but not instantaneous, so using them less often can improve computation time. We find that it is much more efficient to calculate the intensity as a function of planet position (with given limb-darkening coefficients and radius ratio) with a sampling of about 200 positions from 0 to $\frac{R_*+R_p}{R_*}$. Fast cubic interpolation is used to calculate the intensity at all the thousands (or more) of planet positions needed for the observations. One could imagine precomputing all the intensity calculations and then using a large multi-dimensional look-up table to calculate photometry, but we concluded that the interpolation method we employed is sufficient.

The model flux is calculated every $\sim$100-1000 seconds (the time step of the integrator) and then an integration of the cubic interpolate is used to calculate the intensity given the beginning and end times of each observation \citep{2010arXiv1004.3741K}. Note that this correctly accounts for flux changes that may happen within a single integration (\emph{Kepler} long cadence observations which have 30 minute long integrations) and can also account for any particular duty cycle (including aperiodic observations, closed shutter time, read-out time, etc.). We have verified that the timestep size and interpolation step sizes are small enough to not affect the light curve at the 1 part per million level or (usually) less in a variety of circumstances. The photometry and the interpolation is only performed near transits and eclipses to speed up the calculations. 

The flux diminutions due to all the planets is added together in order to calculate the model brightness of the system at the observed time.  (See Section \ref{mutev} below for a discussion of what happens when the planets overlap each other during transit.)  We are also assuming that slow variability due to the star or other data reduction trends on timescales longer than the transits have been successfully removed. Our method, like other methods, also assumes there are no significant changes in brightness during a single photometry calculation of $\sim$100 seconds. This is true for transit and eclipse light curves, but can be more complicated for certain kinds of fast mutual events. For this reason, mutual events are not typically included in our photometric calculations. 

When fitting data, the residuals of the model from the data are computed, squared, and summed to calculate $\chi^2$ in the usual fashion. Data far outside of transits or eclipses is not used in the calculation of $\chi^2$. For $\sim$100-day long simulated \emph{Kepler} photometry at short cadence with a precise tolerance, a single $\chi^2$ computation takes approximately 1 second. To estimate the most likely values for our parameters now requires a minimization of $\chi^2$. This highly non-linear multi-dimensional
parameter estimation is difficult, but is assisted significantly by the fact that the independent Keplerian fits to the transit
photometry and RV data give an excellent first approximation of most of the fitted parameters. 

\begin{figure}						 
\begin{center}
\includegraphics[width=\columnwidth,trim=0in 5in 0in 1in]{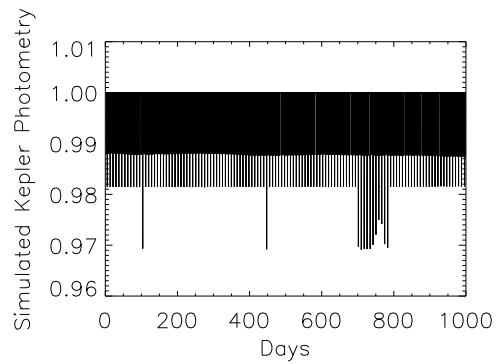}
	\caption{\label{fig:jjres} Simulated Multi-Transiting System showing TTVs. Using the numerical-photometric model described in the text, the light curve shown represents two hot Jupiters in the 2:1 resonance with periods of approximately 3 and 6 days and near central impact parameters. This model includes self-consistently the three-body interactions, light travel-time corrections, and the acceleration of the star from one planet during the transit of another planet, which is not entirely captured in models that use only TTVs and TDVs (see text). The light curve covers 1000 days of simulated \emph{Kepler} photometry at 30 minute (``long'') cadence with no noise added. The two planets were given different depths; the outer planet (with less frequent transits) is clearly larger. Occasionally both planets transits the star at the same time, these double transits have a combined depth greater than either planet individually; mutual events (described in $\S$\ref{mutev} and Figure \ref{fig:odt}) are not included in this model. The first two double transits happen coincidentally and such events would repeat about every year in the purely Keplerian case. The train of double transits around day 750 is caused by planet-planet perturbations and changes in the nominal orbital frequencies and demonstrates that transit timing variations are included in the model. 
         }\end{center}
\end{figure}

An example of some of the simulated photometry from this model is given in Figure \ref{fig:jjres}. Using this model to fit simulated 
photometry, we have drawn the following conclusions. 
\begin{itemize}
\item The global parameter space is covered with many disconnected local minima. We have developed an MCMC-like code 
(based on concepts in \citet{2005AJ....129.1706F} and \citet{2006ApJ...642..505F}) that is coupled with a Levenburg-Marquardt minimization that is run 
every $\sim$50 links (but this is not used to update the MCMC chain as this would violate the Metropolis-Hasting 
algorithm). This combination of a local and global minimizer to explore parameter space and find the best minima is also employed by \citet{2010arXiv1005.5396M}. When started near the global minimum, this code is successful at finding good solutions. For real data, it 
will be  difficult, if not impossible, to be sure that the code has settled in the best global minimum, although reduced $\chi^2$ values near 1 with appropriate errors are a good sign. Estimating 
reasonable errors on the system parameters is also very difficult, especially with limited computational time.

\item The detectability and precision with which parameters are estimated is in accordance with the expected results. 
For examples, in systems where dynamical interactions are important and where TTVs would be observed, we recover the 
correct value of the mass ratios of both planets and we determine the semi-major axes very precisely (as seen also in 
Nesvorny \& Morbidelli 2009). If combined with precise radial velocities, as in Meschiari \& Laughlin 2010, one could 
measure dynamical masses of each of the objects.

\item $\S$\ref{dynmass} discussed the possibility of measuring dynamical masses from photometry alone by observing the 
light-travel time effect. Even with 1000-days of \emph{Kepler} short-cadence photometry, this will not be possible in most systems. 
In a system with two transiting interacting hot Jupiters, the true mass of the system was only constrained to within a 
factor of $\sim$2 with 100-days of data. This implies that (typical) ground-based follow-up observations will be important for understanding multi-transiting systems.

\item Since the two planets generally cross the star along two different chords, one might imagine that 
multi-transiting systems are efficient at breaking limb-darkening parameters. Indeed, this is a major motivation for 
creating a single model that incorporates data for both planets simultaneously. Even so, we found that at the precision 
expected from \emph{Kepler} even on relatively faint stars (1000 ppm/min, V$\simeq$14), the limb darkening parameters are 
already measured by 
the largest planet and the addition of another planet does not substantially improve the measurement of limb darkening. 
However, this conclusion was not tested in detail.

\end{itemize}

\section{Other Techniques for Studying Multi-Transiting Systems}
\label{othersec}

\subsection{Mutual Events}
\label{mutev}

Systems with more than one planet have the unique possibility of photometric mutual events, where the objects shadow or occult one
another as seen from Earth. The primary transit and secondary eclipse are well-known versions of these mutual events, since these ``star-planet'' mutual events are
the only phenomena relevant for single planets. When additional bodies are added, however, other mutual event phenomena are possible.
In multi-transiting systems, the likelihood of observing and correctly interpreting these mutual events is much higher than in systems
where one or no planets transit. Furthermore, if they can be observed, mutual events provide unique constraints on the orbital
parameters of the planets involved and are worth investigating in detail.

As in $\S$\ref{term} above, we begin by being clear with terminology. In a multi-transiting system, when two planets transit 
at the same time, this is called a double transit \citep[following, e.g.,][]{2010EAS....42...39H}. If a mutual event between the planets 
occurs during the transit, we call this an overlapping double transit, which leads to a brightening in the light curve 
as described below. When two planets cross in front of one another outside of transit, we call this a 
planet-planet occultation. Planet-planet shadowing occurs when an inner planet passes between an outer planet and the 
star. Finally, following terminology from the solar system, a mutual event between three objects in an extra-solar 
system could be called a syzygy. Other than the overlapping double transit case (a star-planet-planet syzygy), we do 
not consider other exo-syzygies here. The systems of S10 do not show any double transits in the 43 days of data shown.

Mutual events are very powerful in constraining the orbital and physical properties of the planets. When a mutual event 
occurs, it is like a precise differential astrometric measurement and one can know almost exactly where both of the 
participating planets are at a specific time. In particular, we find that mutual events are excellent at measuring the 
mutual nodal angle, and hence the true mutual inclination, between the participating planets (see discussion below).

In systems were many mutual events can be observed, since these events involve intersections of comparatively small 
planets, they result in precise requirements that must be correctly reproduced in a viable model. One can imagine 
tracking ``mutual event timing (and duration) variations'' which are even more sensitive to small orbital perturbations 
than transit timing variations and consequently provide even stronger constraints on potential perturbations.

We have considered all the possible mutual events between a star and two planets and here describe the possible mutual 
event phenomena in order of detectability: overlapping double transits, planet-planet occultations, and planet-planet 
shadowings. Note that these planet-planet mutual events are intrinsically more rare and shorter-duration than 
star-planet transits since planets are much smaller than stars (and they are not in orbit around one another).

\subsubsection{Overlapping Double Transit}
\label{mutevodt}
If two planets overlap while they are in transit, then the photometric diminution is no longer the sum of two individual diminutions due to
the two planets. The shape of the covered part of the star deviates significantly from two circles and standard light curve calculations can no
longer be used. Even when both planets are known to transit, the probability of such an event depends on the mutual inclination and
node of the planets and these events can occur often or never, even if (non-overlapping) double transits happen regularly. Even if the planes of the planets are properly aligned, the phases of
the planets often make this a rare occurrence in time (assuming resonances are not in play).

\begin{figure}						 
\begin{center}
\includegraphics[width=\columnwidth,trim=0in 0in 0in 0in]{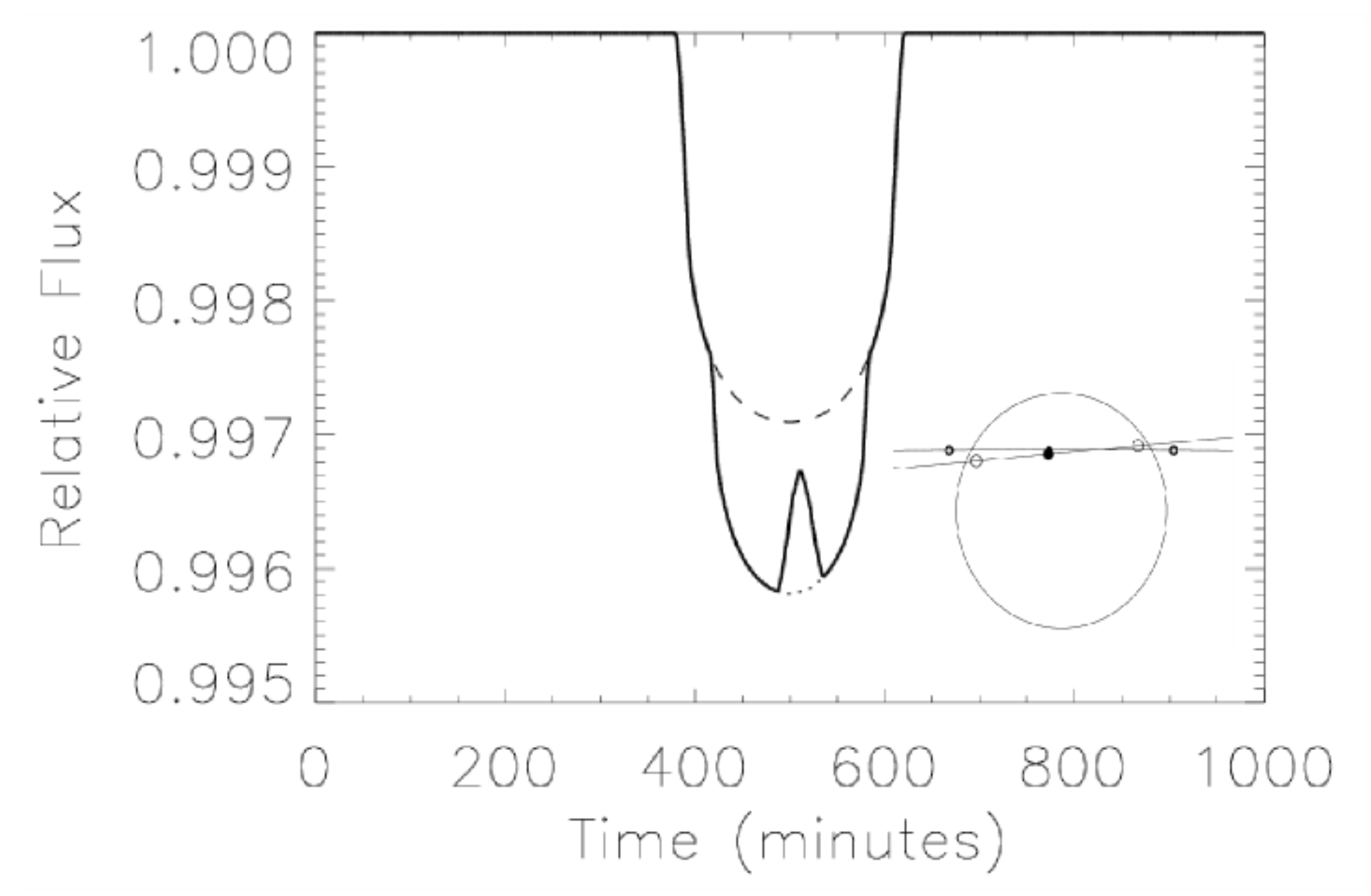}
	\caption{\label{fig:odt} Simulated Overlapping Double Transit. If two transiting planets were to overlap while 
crossing the stellar disk, a mutual event we call an overlapping double transit, they would cover a smaller area of the 
star and a brightening would result. This is shown in the above figure with a simulated mutual event, the observed 
photometry is given by the solid line and would have a distinct ``W'' shape. The dashed line shows what the transit of 
the outer planet would look like alone and the dotted line (mostly concurrent with the solid line) shows what the 
double transit would look like without a mutual event. The difference is a significant brightening whose signal is 
almost as big as the inner planet transit, since the overlap is nearly complete. Inset is a picture showing the two 
orbital paths (straight lines), their mutual inclination of 5 degrees, and three positions of each planet to show which 
orbital path corresponds to which object (they are moving from right to left). The filled figure is the coverage of the 
two planets at the moment of closest approach. Although these are only Neptune size planets, a mutual inclination 
different by only a few degrees would have a detectably different signature by \emph{Kepler} even around a faint star. Mutual 
events are unique and powerful ways of constraining the true mutual inclination, as described in the text. An animation of an overlapping double transit is available for download at \texttt{http://www.cfa.harvard.edu/$\sim$dragozzi/meanim.gif}.
         }\end{center}
\end{figure}

Even so, when an overlapping double transit does occur, it can have easily detectable photometric signal. As mentioned above, early observations by
HST of TrES-1 saw a bump in the light curve that was thought to perhaps be due to a overlapping double transit \citep{2009A&A...494..391R}, but later analysis \citep{2009ApJ...701..756D} showed that a starspot crossing was a much more likely conclusion (see also $\S$\ref{mutevoth} below). When the orbits of both planets are known
well, and given the exquisite knowledge of the periods and phases of high signal-to-noise transiting planets, there will be much less
confusion on this point. Overlapping double transits are also the easiest mutual event to identify: they can only occur during a double transit and are characterized (in the best cases) by an unmistakable brightening (Figure \ref{fig:odt}). Planets crossing over starspots can also cause minor brightenings, but the overlapping double transit usually has enough difference in amplitude and duration that we find it unlikely that a starspot crossing will be confused with a mutual event, especially if the event is not grazing.

If these events happened over a source of uniform brightness, the expected lightcurve could be generated analytically. 
As the star is limb-darkened with both planets moving across it and overlapping, we find that a numerical technique for 
modeling the light curve is required. Since these events are generally rare, this is not computationally costly overall 
and such a model can be folded into the numerical-photometric model described above. To correct the light curve created 
from the sum of the two individual light curves requires adding back in the flux that was doubly subtracted from the 
region of overlap between the two planets. We have generated an efficient method for calculating this light curve 
(similar to Barnes \& Fortney 2003) and tested it with Monte Carlo integrations (similar to Carter \& Winn 2009). 
Resulting light curves of overlapping double transits are given in Figure \ref{fig:odt}\footnote{An animation of an overlapping 
double transit is available for download at \texttt{http://www.cfa.harvard.edu/$\sim$dragozzi/meanim.gif}.}.

We have modeled the case of an overlapping double transit for two Neptune-size planets in circular orbits around a 
relatively faint \emph{Kepler} star ($V \simeq 14$) and find that light curves for mutual inclinations that differ by only a 
few degrees are distinguishable. Indeed, since both planets are transiting the star, all the orbital parameters are 
relatively well known and the only parameter with significant freedom is $\Omega_1-\Omega_2$, the mutual nodal angle; 
since this parameter strongly affects the timing and shape of mutual events, it is not surprising that these events are 
very good at determining $\Omega_1-\Omega_2$ without degeneracy. Other parameters, like the eccentricity, can also be 
constrained as they also affect both the exact positions and exact speeds, which are measured precisely by mutual 
events.

A simple geometric argument can explain why this is the case. Consider the detection of an overlapping double transit, 
where $\Omega_1$ is set to 0$^{\circ}$ without loss of generality. This means that the inner planet traverses an 
essentially horizontal line across the star. The impact parameter is known and the time of the mutual event gives the 
exact orbital phase: from the remainder of the transit light curve, this phase is directly related to the planet's 
position relative to the center of the star and the exact position of the inner planet is known (modulo unimportant 
rotation in the plane of the sky). The second planet is located in the same position and, from its impact parameter and 
phase, is a specific distance from the limb of the star, i.e., the locus of possible limb crossings is a circle of 
fixed radius (since its orbit can be oriented in any direction, not just horizontally). This circle intersects the 
actual limb of the star in one or two locations, immediately limiting the possible trajectories to fewer than two 
possibilities, which translates into a two-way degeneracy in the true mutual inclination. Including information of the 
shape of the mutual event light curve is generally sufficient to break this degeneracy; an extreme example of this 
occurs when one planet is prograde and the other is retrograde, where the mutual event will be much shorter than in a 
system where both planets orbit in the same direction, due to the difference in apparent relative velocity. 
Observable out-of-transit planet-planet occultations have a similar effect of measuring precise orbital parameters, 
including mutual inclination.

\subsubsection{Planet-Planet Occultation}
\label{mutevppo}
Even in multiple-planet systems that are not transiting, it is possible for the planets to cross one another as seen from the
perspective of Earth. \citet{2007A&A...464.1133C} consider such events (as well as planet-planet shadowings discussed below) for planets
and large moons or binary planets as do \citet{2009PASJ...61L..29S} and \citet{2010arXiv1002.0637S}. These authors discuss such mutual events at length for both optical
and thermal measurements, but do not discuss mutual events between unbound objects (e.g., two objects in independent astrocentric
orbits). 

The frequency of planet-planet occultations type of event depends significantly on the geometry of the system. If the planets are coplanar with each 
other
and at a very low relative inclination to the Earth (e.g. nearly central impact parameter), then occultations could occur very frequently. The probability of such an alignment is about $\frac{R_p}{a_2}$, which is $\lesssim$0.5\% even in the most ideal case. 
On the other extreme, planets in very different planes can have rare and aperiodic events. In the case of circular orbits, the apparent orbit on-the-sky is an ellipse with major axis $a$ and minor axis $a \cos i$, with the minor axis of the ellipse oriented at an angle of $\Omega$ east of astronomical North. Mutual events are only possible if these apparent orbits have at least one point where the minimum distance between them is the sum of the planetary radii. This will be most likely for systems with different values of $\Omega$, i.e., those with at least some mutual inclination. Though non-transiting planets can participate in planet-planet occultations, predicting and understanding these events is far easier in multi-transiting systems.

Planet-planet occultations are very weak, since it is the reflected/emitted photons of the blocked planet that are being lost. They are completely
unobservable in systems where the light from the occulted planet cannot be detected, which limits its applicability to occultations of (large) planets with periods $\lesssim$10 days (where reflected light or thermal emission is detectable at some wavelengths with current techniques). None of the systems in S10 are likely to show planet-planet occultations.
The duration of these events is also much shorter
than the secondary eclipse, the usual source of occulting the planet, since the physical sizes of the objects are so small,
though the velocity of the planets on the sky may in unusual cases extend the length of such events. Even in the most favorable cases,
durations are no longer than $\sim$1 hour. Even with fantastic photometry from \emph{Kepler}, \emph{Spitzer}, or \emph{Hubble} space telescopes, individually detecting these events is difficult
even when the outer planet occults a hot and/or reflective inner planet. However, if such events could be detected (or ruled out) with reasonable signal-to-noise, they would place very tight constraints on
the orientation of the planetary system, since there would be information on the relative astrometric position along a line of sight
different than that connecting the Earth the star (see the geometric argument above). 

Furthermore, in these events, high signal-to-noise observations can even characterize the reflected/emitted light
spatially on the occulted planet, possibly leading to surface maps that are not (mostly) degenerate in latitude like those produced by phase curves and secondary eclipse mapping \citep{2007Natur.447..183K,2007ApJ...664.1199R}. We note that this is also true in overlapping double transits \citep{2009arXiv0912.1133K}, though the signal due to the planetary occultation in that case is much more difficult to pick out. 

It might be possible, in well-characterized systems, to predict such events and observe them in the infrared, where the signal is likely to be higher. The \emph{James Webb Space Telescope} may be able to detect some of these short but highly interesting events.

\subsubsection{Planet-Planet Shadowing}
\label{mutevpps}
The smallest effect we consider is planet-planet shadowing. As planets orbit, the shadow of planet 1 can cross planet 2, reducing the amount of light it reflects; from the perspective of outer planet, it is observing a transit of inner planet. If the two planets are distant from one another, the
reflected light from shadowed planet will drop by the depth of the transit, i.e. $\sim$1\% if the inner planet is Jupiter-like. Since the reflected light signal is so faint anyway ($10^{-4}$ at best), this becomes a part per million signal that is very small. You do get a geometric enhancement of about $1 + \frac{a_1}{a_2-a_1}$, but even in the closest orbits, this only amounts to an enhancement of a factor of a few. This effect is maximized for ``double planets'' as has been discussed in this context by \citet{2007A&A...464.1133C}, \citet{2009PASJ...61L..29S}, and \citet{2010arXiv1002.0637S}. 

On the other hand, planet-planet shadowings are likely to be present in systems with low mutual inclinations, they repeat at the potentially-frequent synodic period of the planets, and the duration of such events
can be relatively long. So, it might not be impossible to detect this effect and it could even give constraints on the orbital properties of the planets. However, since the signal is so small, observing planet-planet shadowing may not increase our knowledge of the system if both planets are transiting anyway, though it may be a systematic source of small residuals. Modeling this effect
requires detailed knowledge of the reflected light curve of the shadowed outer planet and would require correct modeling 
of other effects (such as variability due to starspots or planetary weather). 

\subsubsection{Other Effects Related to Mutual Events}
\label{mutevoth}
Mutual events may be relatively rare; very rough estimates indicate that \emph{Kepler} should (serendipitously) observe a few overlapping double-transits, though this likelihood depends strongly on the actual prevalence of multi-transiting systems. It is more common to expect transits of planets across starspots (which have already been observed more than once in known transiting systems). In addition, \emph{Kepler} and \emph{CoRoT} provide long-term light-curve fluctuations that can be used in conjunction with starspot transits, to construct spot models, as in \citet{2010A&A...514A..39H}. In multi-transiting systems, this ability is obviously pronounced, though its use for determining orbital parameters may be limited. Besides an under-constrained spot model, trying to infer the mutual inclination from starspot crossings also requires some knowledge of the stellar spin axis and differential rotation, though perhaps these limitations can be overcome in some systems.

A few very minor effects are worth noting here for completeness. If a close-in planet reflects light back onto the star or otherwise creates a subplanetary hotspot, this can have an effect on the transit light curve of other planets. Similarly ellipsoidal oscillations (and the associated gravity darkening) due to one planet can change the transit properties of another. Very close-in planets with escaping atmospheres could have different mutual event properties in different filters, which might help probe the shape of the escaping atmosphere.

\subsection{Radial Velocities and Multi-Rossiter-McLaughlin}
\label{multiRM}
Though a surprising amount of information about multi-transiting systems can be inferred from photometry alone, radial velocity measurements will improve and confirm other aspects of these systems. As mentioned in $\S$\ref{dynmass}, the combination of TTVs from photometry and RVs gives dynamical masses and absolute radii, which is very exciting. RV observations are also important to search for additional non-transiting planets that do not create observable TTVs. 

Radial velocity observations during transits have particular value: as a transiting planet crosses the parent star, an anomalous radial velocity signal is generated due to the occultation of portions of
the rotationally Doppler shifted star. This effect is known as the Rossiter-McLaughlin effect and is sensitive to the projected angle
between the stellar spin axis and the orbital pole of the planet, called $\lambda$ (equivalent to $-\beta$).
This effect has been measured for several transiting planets (Triaud et al. 2010), with some planets nearly co-planar with their corresponding
stellar equators and many planets showing polar or possibly retrograde orbital paths.

\citet{2009IAUS..253..173F} discussed the possibility of using the RM effect to help constrain the planetary mutual inclination ($\phi$) and \citet{2010arXiv1004.1143S} consider the value of RM on understanding the orbit of transiting exomoons. Here we expand on this discussion to be more specific about the importance of ``multi-RM'' observations. 

Note that the orbital normal to each planet \emph{and} the spin vector of
the star all can be described by a single unit vector or by 3 Euler
angles, assuming that the starting vector is pointing in the
$z$-direction. Using $\omega_*$, $i_*$, and $\Omega_*$ as Euler angles in the standard way for three-dimensional orbit
description proscribe the stellar spin vector as well. (For rotational axes, $\omega_*$ is not
important, unless you're interested in the rotational longitude at epoch.)

Using this notation, the true angle between the star's spin axis and a planet's orbital axis, $\psi$, can be written as
\be
\cos \psi = \cos i_* \cos i_p + \sin i_* \sin i_p \cos (\Omega_*-\Omega_p)
\ee
and comparing this to the definition of $\lambda$ from \citet{2009ApJ...696.1230F}, we find that $\lambda \equiv \Omega_* - \Omega_p$, i.e., the difference in nodal angles. Whereas in the case of transiting planet orbits, where both inclinations are known, but the relative nodal angle is not, the Rossiter-McLaughlin effect is sensitive to the relative on-the-sky alignment angle between the star and planet, but the stellar inclination to the line-of-sight, $i_*$ is not determined. 

This difference can be exploited in multi-transiting systems and it is trivial to show that $\Omega_1 - \Omega_2 = |\lambda_1 \pm \lambda_2|$. This means that if two transiting planets have measured Rossiter-McLaughlin angles, the true mutual inclination, $\phi$, is determined to a two-way degeneracy. If one of $\lambda_1$ or $\lambda_2$ is small (i.e. aligned with the star) then the degeneracy is weak and mutual inclination can be determined well, assuming that the errors on the values of $\lambda$ are small (often the errors are substantial, especially on fainter stars). In a triple-transiting system, measurement of all three values of $\lambda$ lift the degeneracy completely, and all the mutual inclinations are directly measured.

This is a promising result, since the other major methods for measuring true mutual inclination -- TTVs and mutual 
events -- will not be possible for all systems and/or may require many years of observations. Multi-RM is 
typically possible in 
all multi-transiting systems and is limited only by the brightness of the star, stellar type, stellar rotational velocity, the sizes of 
the planets, and other observational constraints. This suggests that a campaign of multi-RM measurements can be used 
to probe the mutual inclination distribution of planets and provide important insights into their formation and 
evolution ($\S$\ref{mutinc}). Such a campaign would also, clearly, probe projected star-planet alignment for each of these 
planets, which is also relevant to their formation and evolution and should provide useful clues to understanding 
various properties of star-planet alignment in general (Triaud et al. 2010; Winn et al. 2010, submitted).

Finally, since two different planets cut different chords across the star, it may be possible to break some of the degeneracies normally associated with the RM effect, though the effects of differential rotation should be considered in this regard. For example, the (latitudinal) motion of star spots depends on the unknown $i_*$, and the timing of spot crossings can potentially constrain this parameter \citep{2008ApJ...682..586M}; the likelihood of useful spot crossings seems much higher in multi-transiting systems and \emph{Kepler} provides long-term light-curve fluctuations that can be used to constrain spot models ($\S$\ref{mutevoth}). 

\section{Conclusions}
\label{concl}

The announcement of candidate multi-transiting systems by S10 ushers in a new era for transiting exoplanets. Besides great potential for informing theories of planetary formation, planetary interiors, orbital migration, large-scale scattering, and other effects, multi-transiting systems are among the best for understanding the question of planetary habitability. The radii of potentially habitable super-Earths are measured photometrically, and their masses can sometimes also be measured photometrically by TTVs induced on other transiting planets, which is especially valuable in the case where RV measurements are not feasible \citep{2010EAS....42...39H}. Only with good estimates for masses and radii can habitability be seriously assessed. Furthermore, the orbital architecture of the entire system is much better understood if more than one planet is transiting the star, which is important for such questions as long term stability and potential climate cycles.

We have attempted a systematic investigation of the value of systems with multiple transiting planets. Many of 
the basic ideas have been presented by other authors \citep[especially][]{2004ESASP.538..407S,2009IAUS..253..173F,2010EAS....42...39H,2010arXiv1006.2763S}, but 
we have expanded their results in a way that will be helpful for maximizing the value of the observations. This has 
resulted in the following conclusions.  
\begin{enumerate}
\item Based on geometric constraints and RV observations, multi-transiting systems are likely to be composed of close-in multiple Neptune systems with low mutual inclinations. While the frequency of these systems remains to be seen, the identification of at least five good multi-transiting candidate systems by S10 using only 43 days of data implies that \emph{Kepler} will discover several multi-transiting systems.

\item Transiting planets in general are useful for obtaining important physical parameters about distant exoplanets, including mass, radius, and density. Multi-transiting systems do this efficiently for more than one planet in the same system. Comparative planetology between these planets is more powerful than between planets in disparate systems.

\item Multiple planet systems in general are useful for understanding the formation and evolution of planetary systems. Multi-transiting systems are multiple systems that will be very well characterized, with true masses and other orbital parameters (such as resonance occupation) precisely measured, along with a well-determined orbital architecture (although potentially missing some non-transiting planets that are too small to generate observable TTVs or RVs).

\item Multi-transiting systems can be mimicked by false positives, and each candidate planet should be confirmed 
separately, though it is increasingly difficult to develop realistic astrophysical models for systems with many planet 
candidates. Consistent inferred stellar properties (i.e., density and limb darkening) are important to check, but are 
not 
generally sufficient to prove the planetary hypothesis. Once the candidates can be shown to orbit the same star, (e.g.,  
through the identification of robust and self-consistent TTV signals), it is generally straightforward to prove that 
their masses must be small (e.g., through TTVs (or lack thereof), orbital stability, or other theoretical grounds).

\item Multi-transiting systems that show TTVs are much easier to interpret since the size, period, and phase of the 
perturbing planet are known. Combination of TTVs with RV measurements can allow for the determination of dynamical 
masses and absolute radii for each of the objects in the system \citep{2005MNRAS.359..567A}. In the best cases, dynamical mass 
limits can be placed on these systems through photometry alone, if light-travel time or photometric Doppler boosting 
\citep{2003ApJ...588L.117L} are detected, but basic ground-based follow-up will always be justified.

\item Combining the inclination distribution of exoplanets with other properties, such as eccentricity and star-planet 
alignment distributions, will be extremely valuable for interpreting the formation and evolution of individual multiple 
systems and planetary systems in general. 
Although the true mutual inclination is not determined directly in multi-transiting systems, there are several useful 
techniques that can be applied to these systems that can constrain this valuable parameter including TTVs, exoplanet 
mutual events, and multiple Rossiter-McLaughlin measurements. Multi-RM observations constrain the true inclination with 
comparable errors as the star-planet alignment angle, $\lambda$, though in cases where two planets are mis-aligned 
there is a two-way degeneracy. Multi-transiting systems offer the best hope for probing the inclination distribution of 
planetary systems.

\item We have developed a full numerical-photometric model that accounts for all the important effects of interacting 
planetary systems and photometric diminutions. In such a model, the acceleration of the star due to a perturbing planet 
can be measured and characterized; this is not entirely captured in models that only fit TTVs and TDVs. The model also 
exploits 
the fact that both planets transit the same star (with the same radius and limb darkening), and accurately accounts for 
light-travel time offsets. Due to computational limitations, this model is best used as the final step in determining 
orbital and physical parameters in multi-transiting systems.

\item Mutual events between exoplanets, including overlapping double-transits, planet-planet occultations, and 
planet-planet shadowings, are very valuable, especially for measuring the mutual inclination between two 
transiting planets. The 
strongest signal is given by overlapping double-transits, for which we have developed an efficient model. Though not as 
precisely constrained as mutual events, starspot crossings may also have some value as they can be more common. 
Finally, planet-planet occultations may be valuable for probing the surface brightness distribution of exoplanets in a 
unique way.

\end{enumerate}

We have seen from the study of stars that eclipsing double-lined spectroscopic binaries are the observational 
foundation on which much of stellar astrophysics is based. We propose that stars with multiple transiting planets will 
take a similar role in characterizing extra-solar planets and shaping our models of the formation and evolution of 
planetary systems. For the reasons outlined in this work, multi-transiting systems are the most information-rich 
planetary systems outside our solar system around main sequence stars\footnote{Pulsar timing can measure orbital 
characteristics of ``pulsar planets'' much more precisely than photometric measurements \citep[e.g.,][]{1994Sci...264..538W}. However, the radii of pulsar 
planets are not measurable and their orbital characteristics are difficult to connect to the formation and evolution of 
the original planetary systems.}, and should receive special attention.

\begin{acknowledgments}

We thank Dan Fabrycky, Eric Ford, Dave Latham, Dimitar Sasselov, Jason Steffen and others for comments and suggestions. 

\end{acknowledgments}


\bibliographystyle{apj}
\bibliography{exo}

\end{document}